\documentclass[final,5p,times,onecolumn]{elsarticle}
\usepackage{amssymb}
\usepackage{amsmath}
\usepackage{graphicx}
\usepackage{xfrac}
\usepackage{hyperref}
\usepackage{lineno}
\usepackage{xcolor}

\hypersetup{
	colorlinks,
	linkcolor={black},
	citecolor={blue!50!black},
	urlcolor={blue!80!black}
}
\makeatletter
\def\@author#1{\g@addto@macro\elsauthors{\normalsize%
    \def\baselinestretch{1}%
    \upshape\authorsep#1\unskip\textsuperscript{%
      \ifx\@fnmark\@empty\else\unskip\sep\@fnmark\let\sep=,\fi
      \ifx\@corref\@empty\else\unskip\sep\@corref\let\sep=,\fi
      }%
    \def\authorsep{\unskip,\space}%
    \global\let\@fnmark\@empty
    \global\let\@corref\@empty  
    \global\let\sep\@empty}%
    \@eadauthor={#1}
}
\makeatother

\journal{Nuclear Instruments and Methods in Research A}

\begin{document}
\begin{frontmatter}

\author{J.D.~Maxwell\corref{cor1}}
\ead{jmaxwell@jlab.org}
\author{C.S.~Epstein}
\author{R.G.~Milner}
\author{M.~Musgrave}

\address{Laboratory for Nuclear Science, Massachusetts Institute of Technology, Cambridge, MA USA}
\cortext[cor1]{Corresponding author, current affiliation Jefferson Lab}

\author{J.~Alessi, G.~Atoian, E.~Beebe, A.~Pikin, J.~Ritter, A.~Zelenski}

\address{Collider-Accelerator Department, Brookhaven National Laboratory, Upton, NY USA}

\title{Enhanced Polarization of Low Pressure $^3$He through Metastability Exchange Optical Pumping at High Field}

\begin{abstract}
We report high steady-state nuclear polarization of 1\,torr $^3$He gas nuclei via metastability exchange optical pumping at magnetic fields above 2\,T. The introduction of highly polarized $^3$He gas into Brookhaven's Electron Beam Ion Source would enable a new, polarized $^3$He ion source for use at the Relativistic Heavy Ion Collider and a future Electron--Ion Collider facility. By adapting recent developments in high field metastability exchange optical pumping for higher pressure gas, we have successfully polarized 1\,torr $^3$He sealed cells in the EBIS solenoid. Through careful manipulation of the RF discharge parameters, polarizations above 80\% were attained at 2, 3 and 4\,T, with 89\% being reached at 3\,T with a 664\,s relaxation time. 
\end{abstract}

\begin{keyword}
$^3$He polarization \sep metastability exchange optical pumping \sep high magnetic field


\end{keyword}

\end{frontmatter}


\section{Introduction}
Metastability exchange optical pumping (MEOP) allows enhanced polarization of $^3$He gas nuclei in a uniform magnetic field using circularly polarized light \cite{colgrove}. An RF discharge in the gas promotes a small fraction of the atoms into the 2$^3$S$_1$ metastable state. Transitions from the 2$^3$S$_1$ into the 2$^3$P$_0$ states are driven using 1083\,nm laser light, which will change the magnetic quantum number by $\pm 1$ depending on the circular polarization of the light. The polarization induced in the metastable population is then transferred into the ground state population via metastability exchange collisions. Polarizations of 85\% have been reached in a 0.3 torr sealed cell under a 10\,MHz RF discharge and 4.5\,W of pumping laser power at 1.2\,mT~\cite{gentilemck}. The excellent review paper by Gentile, Nacher, Saam and Walker~\cite{gentile} provides in-depth background on this technique and its capabilities.

Our effort to introduce a polarized $^3$He ion beam to Brookhaven National Laboratory's Relativistic Heavy Ion Collider (RHIC) utilizes MEOP to provide polarized gas for ionization and extraction by BNL's electron beam ion source (EBIS)~\cite{alessi,maxwell2}. While MEOP techniques have traditionally offered high polarization in only a limited pressure range---typically around 1 torr---the high rate of polarization and the ability to polarize without impurity from a quench gas or alkali vapor  makes MEOP attractive for this application. 

Traditional MEOP is performed in a low magnetic field \textemdash typically below 15\,mT\textemdash{}so our initial $^3$He ion source concept required magnetically shielding the pumping cell from the EBIS 5\,T solenoid's nearby stray field. Polarized gas would then be transferred from the 1 torr, 3\,mT pumping cell, through the stray field of the solenoid, into the $10^{-7}$ torr, 5\,T drift tubes of EBIS. Studies of polarized gas transit through regions of depolarizing field gradients~\cite{maxwell3} indicated that while such a transfer may be feasible, it would require complicated shielding schemes and ultimately some compromise in the achievable final polarization.

Another approach to a MEOP-based ion source with EBIS would be the production of polarized $^3$He gas in a cell near or perhaps within the 5\,T field. At first glance, one might expect the strong magnetic field would reduce the electron--nucleus spin coupling that allows nuclear polarization, making MEOP untenable. However, a group at the Kastler Brossel Laboratory (LKB) in Paris discovered a surprising improvement in MEOP efficiency for high pressure gas at high magnetic field~\cite{Courtade2000}. While they have been primarily driven to improve polarization performance for high gas pressure applications, such as medical imaging, they showed that MEOP techniques can be successfully extended to higher fields. In 2004, they achieved 80\% steady-state polarization at 1 torr and 1.5\,T~\cite{abboud,nikiel}. In 2013, further results at 4.7\,T showed continued success in the refinement of high-field MEOP techniques, particularly at pressures above 20 torr ~\cite{nikiel2}. However, in this study their 1 torr gas cell reached only 32\%. Taken together, these measurements suggest that low pressure pumping cells may not enjoy the same improvement of MEOP efficiency with increasing field that was observed at higher pressure.

To investigate the feasibility of polarizing closer to EBIS, we set out to understand how MEOP efficiency at 1 torr falls off between these two previous measurements at 1.5 and 4.7\,T. Of key interest is the order of magnitude disparity in the polarization decay time measured with plasma discharge by Nikiel \textit{et al}~\cite{nikiel2} between their 24 to 200 torr cells and 1 torr cell. They note that ``the rather low polarization obtained at low gas pressure may be due to unfavorable plasma conditions in the present work.'' Difficulty in lighting and maintaining a plasma discharge can greatly affect the relaxation rate and maximum steady-state polarization, as can the cleanliness of the cell walls, so finding a way to improve the decay time from the 80-150\,s they observed could be a crucial step to attaining significantly higher polarization for our polarized source.

\subsection{MEOP at High Field}
\begin{figure}
	\begin{center}
		\includegraphics[width=\columnwidth]{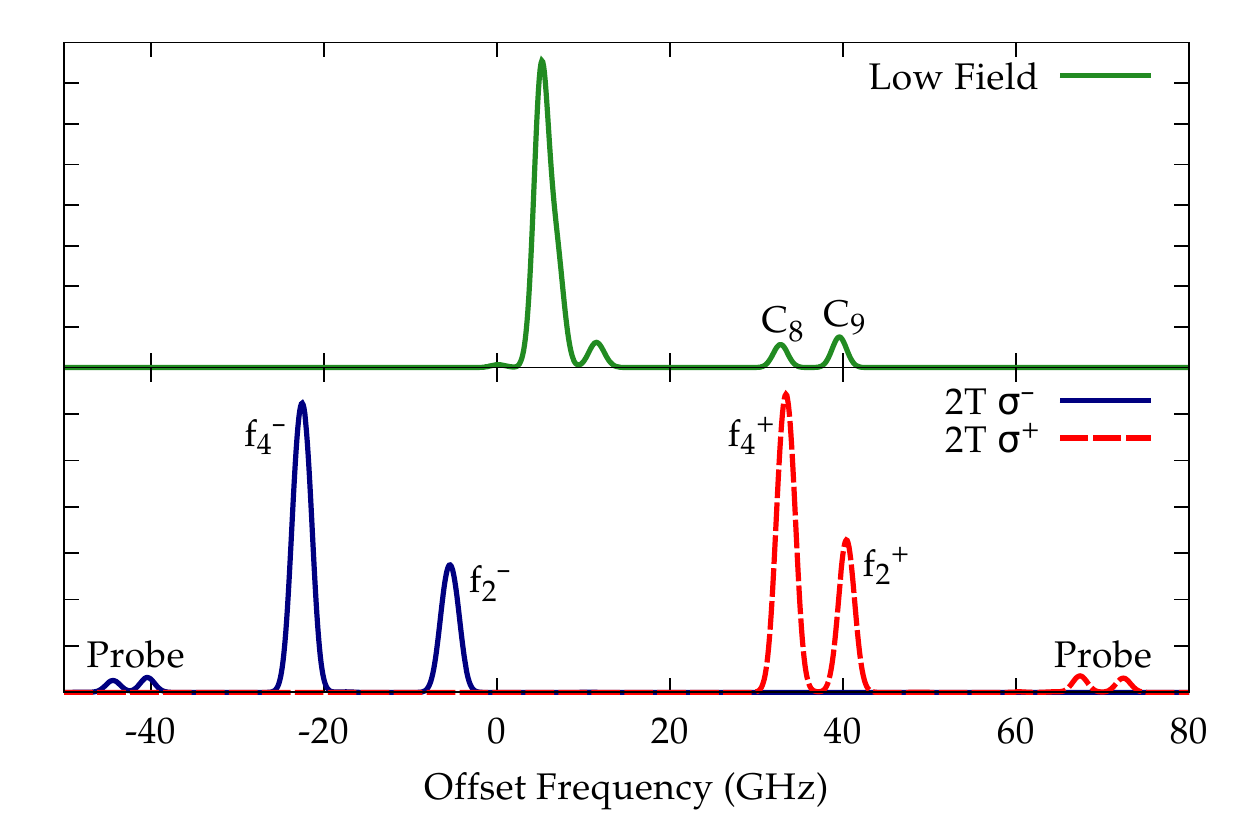}
	\end{center}
	\caption{Absorption spectrum of $^3$He at low magnetic field and 2\,T, showing the separation of degenerate Zeeman states with increasing field. Pumping transitions are labeled by name, as are lines useful for probes.}
	\label{fig:spec}
\end{figure}

In traditional, low-field MEOP applications, the $C_8$ or $C_9$ transitions are used to pump the $2^3S_1$ spin \sfrac{1}{2} and \sfrac{3}{2} states into the $2^3P_0$ state~\cite{gentile}. As the magnetic field increases, the angular momentum structures in the $2^3S$ and $2^3P$ levels shift significantly, breaking the degeneracy of the states and creating new lines in the absorption spectra~\cite{courtade}. Figure \ref{fig:spec} shows absorption lines of $^3$He both at low field and at 2\,T, including Doppler widths for gas at 1\,torr and 300\,K, as calculated using a Fortran routine provided by P.J.~Nacher~\cite{nacheremail}. The lines have separated not only by their energy, but also by the polarization of the light required to drive the transitions~\cite{gentile}. 

The high-field modified levels present new choices for optical pumping schemes. The strongest components in the spectra are labeled as $f_n^\pm$, with sign denoting the circular polarization and the $n$ referring to the number of transitions in the line. These strong transitions offer the most obvious place for optical pumping, with the $\sigma_{-}$ lines having been reported as most efficient~\cite{nikiel2}, particularly at higher pressures. Collision-induced broadening makes the $f_4^+$ and $f_2^+$ peaks more difficult to resolve by a pumping laser as pressure increases, and as they have opposite polarizing actions, this can be problematic~\cite{gentile}. However at low pressure and higher field, these peaks are sufficiently separated to pump individually. Observing the weak, ``probe'' lines in the spectra  with a secondary laser offers a convenient view of the population of the states in the gas without significantly affecting the polarization, as discussed in Section~\ref{sec:polar}.

\section{Apparatus}
\begin{figure}[t]
	\begin{center}
		\includegraphics[width=\columnwidth]{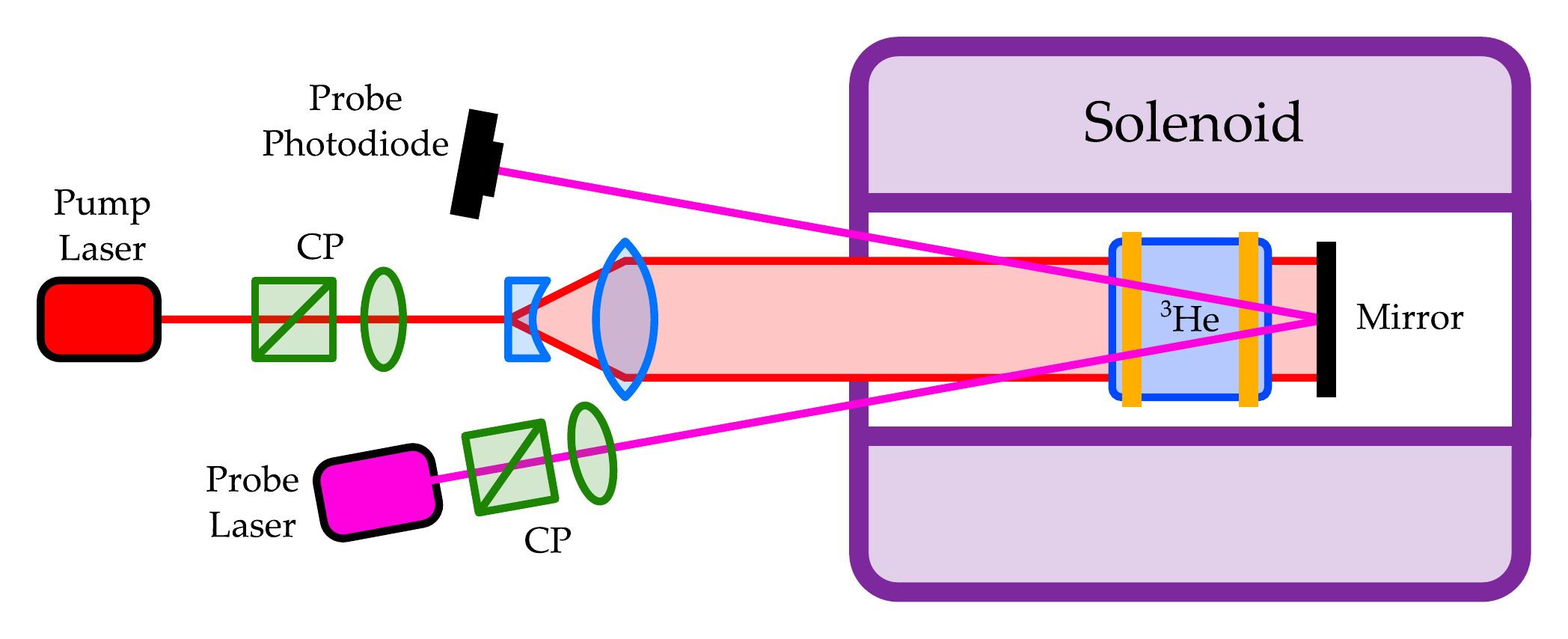}
	\end{center}
	\caption{Diagram of high field polarizing apparatus and probe laser polarimeter.}
	\label{fig:setup}

	\begin{center}
		\includegraphics[width=\columnwidth]{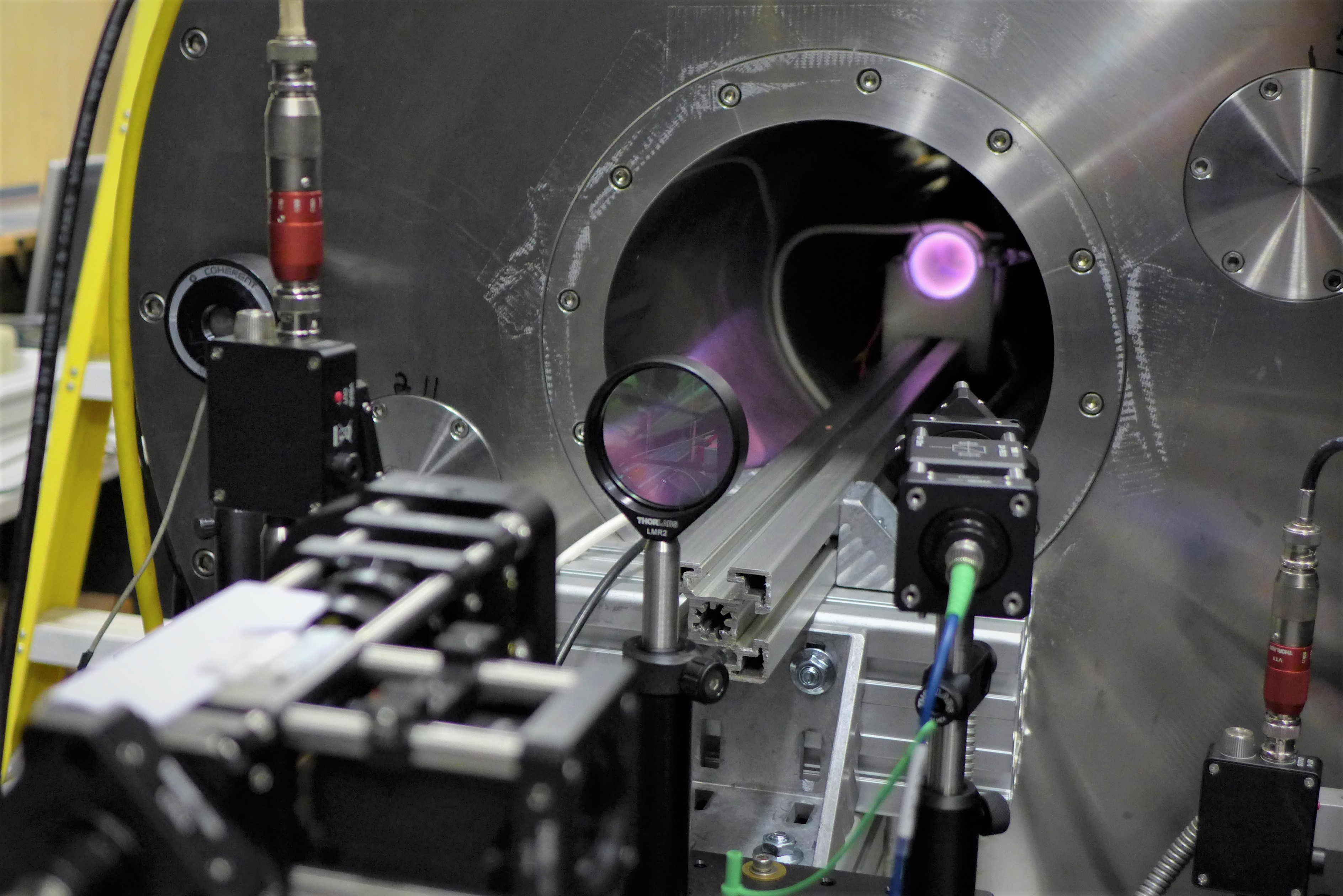}
	\end{center}
	\caption{Photograph of the polarizing apparatus and EBIS spare solenoid warm bore. In the foreground are the pumping laser circular polarization optics. The probe laser fiber enters a circular polarizer on the right, and after passing through the cell the probe light is reflected by a mirror back to a photodiode on the left. The sealed cell is illuminated by the RF discharge plasma in pink; for this photograph it is much brighter than is effective for optical pumping.}
	\label{fig:photo}
\end{figure}

Tests of MEOP at fields up to 5\,T were performed at the Brookhaven Collider-Accelerator Department laboratory using a spare EBIS superconducting solenoid magnet. A diagram overview of the setup, as well as a photograph of the apparatus in the solenoid warm bore are shown in Figures~\ref{fig:setup} and \ref{fig:photo}.  A Keopsys continuous-wave, Ytterbium-fiber laser provides 1083\,nm pumping light at up to 10\,W with a nominal 2\,GHz linewidth, and allows tuning of the wavelength over a 100\,GHz range. A polarization-maintaining optical fiber delivers the light to a polarizing cube to ensure full linear polarization, before the light is circularly polarized using a zero-order, $\lambda/4$ wave plate. The pump light is then expanded and collimated to illuminate the full volume of the gas cell.

To produce the metastable population within the gas, a plasma discharge is induced using radio frequency voltage across electrodes on the exterior of the cell. With increasing magnetic  field, the charged plasma tends to become more intense at the periphery of the cell and thinner in the center, although we did observe strong dependence on the RF frequency and tuning parameters in the spatial distribution of the plasma. Generally a driving frequency around 10\,MHz was effective to produce the discharge, however, frequencies as high as 100\,MHz sometimes provided a more stable discharge. Some dependence on RF frequency is expected~\cite{lorenzon}, but this is largely a factor of finding the most dependable, low-level discharge in given parameters. The discharge was driven by a SRS signal generator, amplified by a 40\,db RF amplifier, and tuned with an MFJ 945E radio transformer before delivery to the cell's electrodes. To first ignite the discharge, a small, high-voltage electric spark was used, after which the discharge was reduced as much as possible. To achieve the highest polarizations, the weakest discharge achievable is most effective, so the RF voltage was reduced either by hand or step-wise by a routine which took input from a photodiode observing the brightness of the cell.

While the final ion source polarizer will require some ingress and egress of gas, for these tests we have used sealed borosilicate glass cells. These cells were 5\,cm in diameter and 5\,cm long, and consisted of 5\,cm optically-flat, Borofloat windows joined by a borosilicate glass tube.  One such sealed cell was on loan from T. Gentile at NIST. Others we produced at MIT Bates through repeated baking under vacuum, purging with clean $^4$He gas, and lighting strong RF-discharges, before they were filled with $^3$He to 1\,torr. The cleanliness of the cell and purity of the gas within can be monitored at different points in the process by alternately observing the spectrum of light emitted during a plasma discharge for the expected helium lines, and sampling the gas make-up with a mass spectrum analyzer.

To locate the absorption spectrum peaks to be used for pumping, the pump laser frequency can be swept as the intensity of the fluorescence light emitted from the cell is monitored. This fluorescence---due to emission from excited states in the discharge---is then directed to a photodiode using a fiber optic line observing the gas cell radially.

Data acquisition and control was performed using a LabView program and a National Instruments USB-6212 multifunction DAQ system. This DAQ was used to provide analog voltage outputs to change the pump and probe laser frequencies and to monitor the analog input voltage from two Thorlabs DET50B photodiodes monitoring the probe laser intensity and fluorescence due to the pump laser light.

\subsection{Polarimetry}
\label{sec:polar}
At low magnetic fields, MEOP enhanced polarization is easily measured via the circular polarization of the emitted 668\,nm discharge light~\cite{pavlovic,maxwell}. As increasing magnetic field weakens the hyperfine coupling efficiency that connects that particular transition to the ground state polarization, this scheme is not tenable above 10\,mT for 1\,torr gas. Directly probing state transitions with laser light offers polarization measurements that are effective at low and high magnetic field~\cite{talbot,suchanek}. At low field, this can be done by probing the $C_8$ line, as the $\sigma_+$ and $\sigma_-$ polarization components each address a single sublevel. Our first measurements with the probe laser were made to confirm agreement with our existing 668\,nm discharge polarimeter at well-known, 3\,mT conditions.

At high magnetic field, the population of two particular $2^3S$ sublevels can be monitored by sweeping the probe laser frequency, providing a measure of the ground state polarization. These sublevels are chosen to avoid the states under active pumping for polarization; for example, if $f_4^+$ is being pumped to drive states $A_3$, $A_4$, $A_5$, and $A_6$, the states designated as ``probe''  for the $\sigma_+$ lines in Figure~\ref{fig:spec} can be used to probe sublevels $A_1$ and $A_2$ (see reference \cite{gentile}, Figure 16 for a detailed diagram of the substates). While these two lines appear weakly in the spectrum, they are well resolved by a probe laser and minimize any impact on the large population changes induced by the pumping laser. At spin-temperature equilibrium, the populations of these probed states, here $a_1$ and $a_2$, will satisfy $a_2/a_1 = e^\beta = (1+M)/(1-M)$. An absolute measure of the nuclear polarization $M$ of the ground states can be formed from the change in the ratio $r=a_2/a_1$ of the absorption signal amplitudes for these sublevels during MEOP, as calibrated by their ratio $r_0$ when not polarized ($M=0$):
\begin{linenomath}\begin{equation}
M = \frac{r/r_0-1}{r/r_0+1}
\label{eq:pol}
\end{equation}\end{linenomath}
Because only ratios of spectral amplitudes are involved, all experimental parameters affecting the absolute signal intensities are canceled out~\cite{suchanek}, making this a robust measurement.

To measure the polarization in our sealed $^3$He cells, we built an optical probe polarimeter in the style of the LKB group~\cite{abboud,nikiel2}, with a few adaptations for our circumstances. As shown in Figure~\ref{fig:setup}, the probe light is circularly polarized by a splitter-cube and $\lambda/4$ wave plate before being directed through the $^3$He cell to a mirror, which reflects the light back through the cell to a photodiode outside the solenoid. To reduce the total power incident on the cell from the probe laser, an iris aperture is used to reject much of the probe light after it has been circularly polarized.

To isolate the absorption spectrum signal from noise in the photodiode and improve the sensitivity of the measurement, a modulation scheme was adopted. While a scheme involving both modulating  the discharge and chopping the probe laser has been utilized by others~\cite{nikiel2}, we found a single modulation of the RF source is sufficient for our needs. The SRS SG382 signal generator, which provides the RF source for our discharge, is amplitude modulated at 1\,kHz at a depth of 50\%. This variation of the discharge intensity induces a synchronous change in the $2^3S$ states, and thus the optical thickness of the gas that is visible via the probe light intensity incident on the photodiode. Instead of a stand-alone lock-in amplifier device, we utilize a tone analysis module in LabView, which acts as a virtual lock-in for our purposes. This module identifies the amplitude of components of a given frequency in the signal, giving us a measure of the probe laser absorption which rejects any signal due to light incident on the photodiode at different frequencies.
  
The probe laser light is provided by a Toptica, 70\,mW, 1083\,nm DFB laser system, which allows modulation of the frequency through control of either the chip temperature or the operating current. For our purposes, we use the temperature modulation to explore the entire range of the probe laser frequency, mapping the absorption spectrum peaks, then use the faster current modulation to sweep the frequency over just the two probe peaks for measurements. By changing the current to perform a frequency sweep, we are at the same time changing the intensity of the probe laser light, albeit by a small amount. Following Nikiel~\cite{nikiel2}, we divide the absorption signal by the average photodiode voltage to remove the change in probe power over the sweep.

\begin{figure}
	\begin{center}
		\includegraphics[width=\columnwidth]{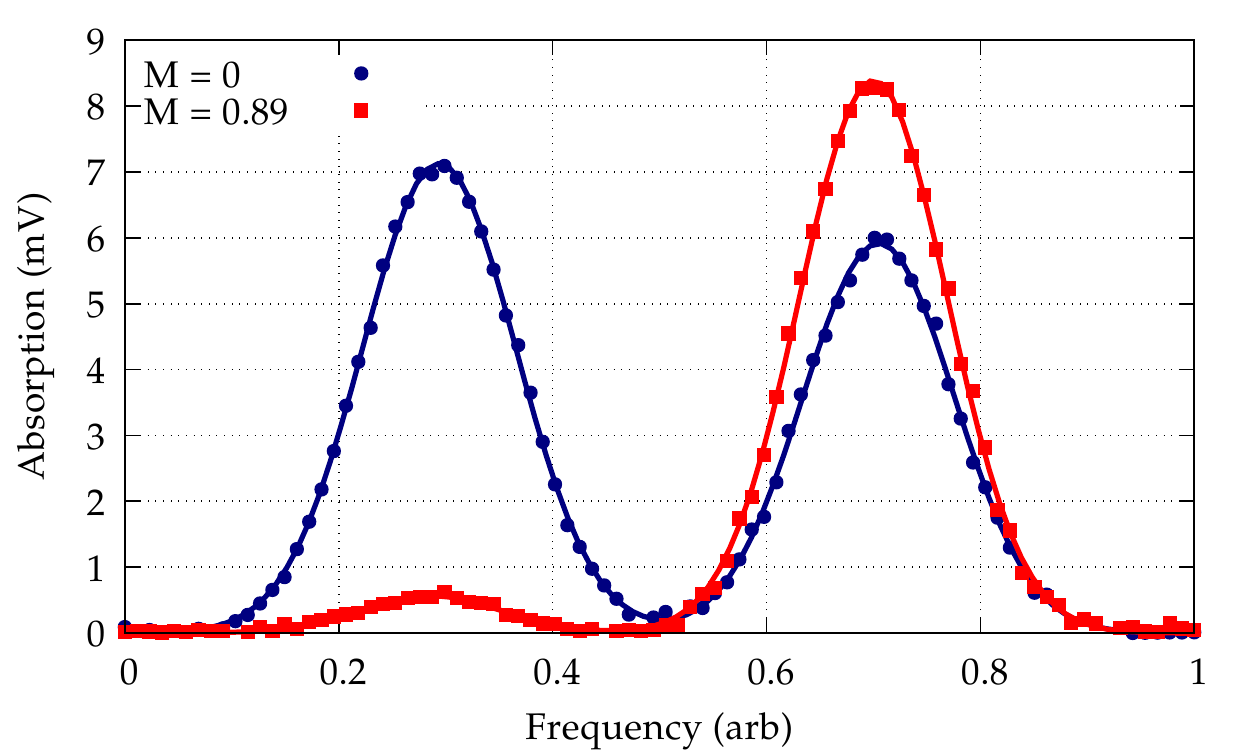}
	\end{center}
	\caption{Example probe laser absorption signals with sample nuclear polarization at 0 and 89\%, using a 1\,torr sealed cell at 3\,T. Both probe transition peaks are visible for each signal, as are the side-by-side Gaussian fits used to extract the peak amplitudes for analysis.}
	\label{fig:example}
\end{figure}

To perform a measurement, first the absorption spectrum of the plasma discharge is mapped through the full range of the probe laser's temperature range. Proceeding from high to low temperature (low to high frequency), a spectrum like that in Figure~\ref{fig:spec} should be seen, either $\sigma_+$ or $\sigma_-$  depending on the orientation of the $\lambda/4$ plate. Next, the temperature is set so that an appropriate sweep in current will cover the frequency range of both probe peaks. Sweeps of probe laser frequency over the two probe peaks proceed continuously during measurements, with each sweep consisting of 90 frequency samples in roughly 6\,s. The absorption peaks, as measured as a voltage from the photodiode, are each fit with a Gaussian function to extract their amplitude, as seen in Figure~\ref{fig:example}.

To calibrate the probe polarimeter, the peak amplitudes are recorded with zero gas polarization. This is generally done before the pump laser is turned on or multiple relaxation times after the pump laser is off. When optical pumping is started, the zero polarization peak ratio, $r_0$, is used with the current peak ratio, $r$, to form polarization $M$ according to Equation~\ref{eq:pol}.
Figure~\ref{fig:example} shows two such probe laser frequency sweeps through the two probe peaks, at zero polarization and at $M=89\%$.

\section{Results}

\begin{figure}
	\begin{center}
		\includegraphics[width=\columnwidth]{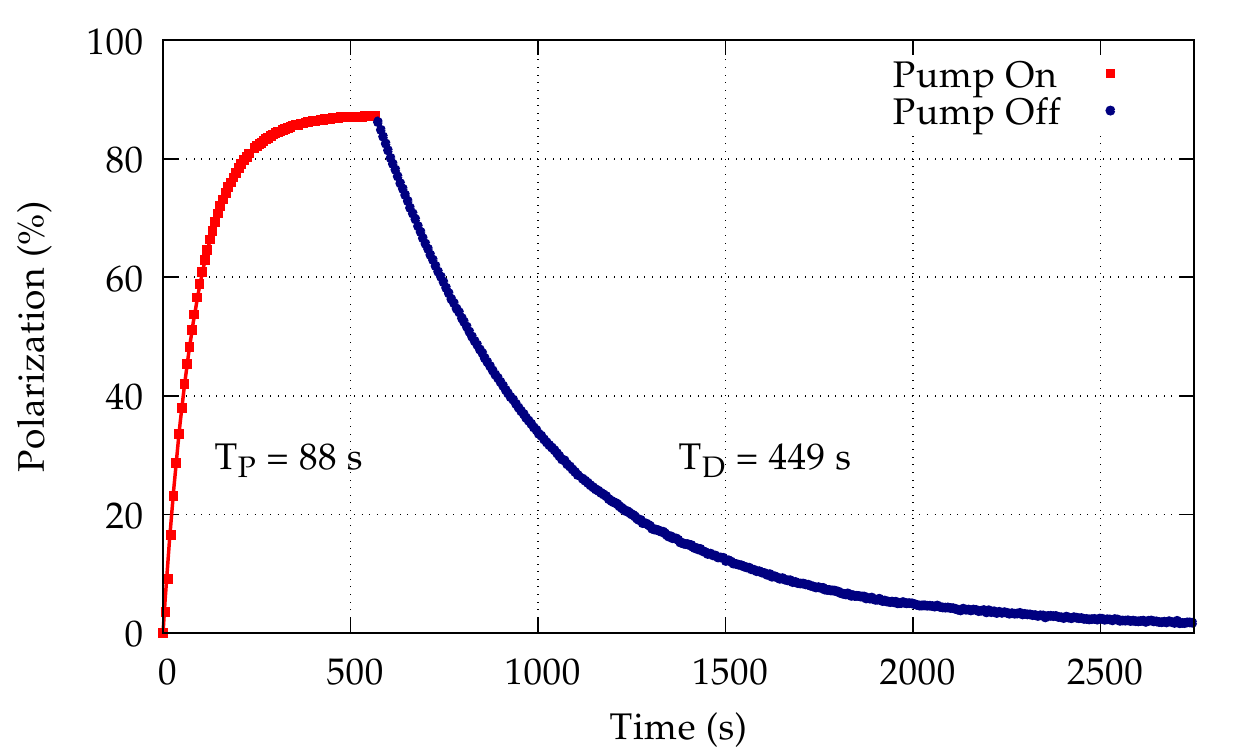}
	\end{center}
	\caption{Typical pump and relaxation cycle at 2\,T, showing exponential pumping build-up time $T_P$ with the optical pumping active, and the relaxation time $T_D$ after the pumping laser is blocked at 560 s.}
	\label{fig:relax}
\end{figure}

Figure~\ref{fig:relax} shows a typical cycle of polarization build-up and relaxation as measured in our 1\,torr cell at 2\,T using the probe laser polarimeter, where the pump light is turned on at 0\,s and blocked at 560\,s. Notable is the purely exponential build-up apparent at high field, compared to the varying polarization build-up rate seen in  low field MEOP~\cite{nikiel2}. The measured relaxation time $T_D$ with the laser blocked, 449\,s in the figure, is dominated by the contribution of the RF discharge and thus is useful as a way to characterize the plasma conditions. With the discharge off, the relaxation time depends on factors such as the field uniformity and wall conditions. While measurement of the polarization with our probe is not possible with discharge off and no metastable states to pump, by making intermittent measurements in short bursts, we measured a discharge-off relaxation time of roughly 10,000 seconds in these conditions.

\begin{figure}
	\begin{center}
		\includegraphics[width=\columnwidth]{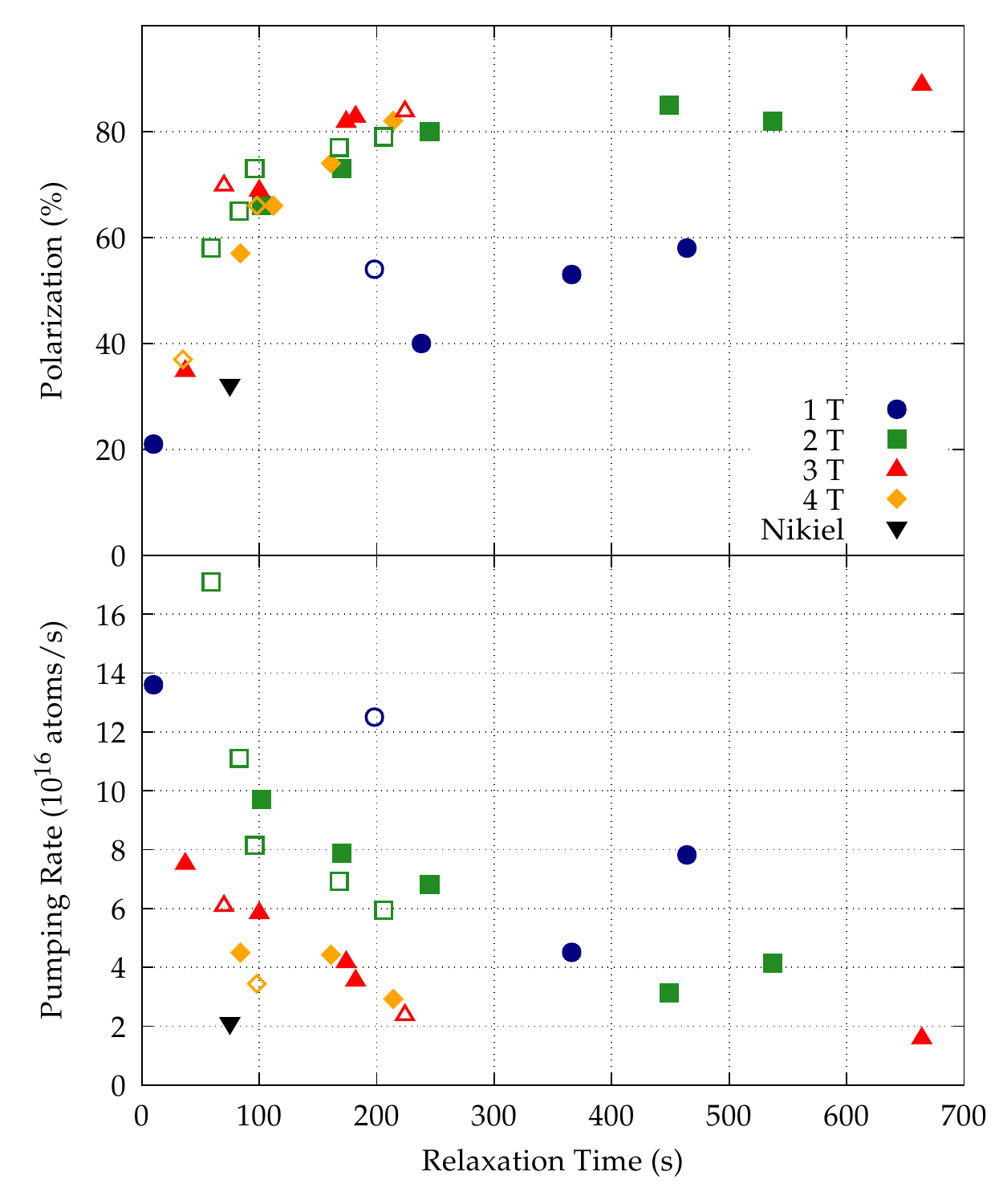}
	\end{center}
	\caption{Steady state polarization achieved (top) and corresponding pumping rate (bottom) for a given relaxation time with the plasma discharge. Different shapes represent four magnetic field settings from 1 to 4\,T. Filled shapes designate measurements on a 1\,torr sealed cell produced a MIT Bates, while open shapes designate those taken on a 1/,torr sealed cell on loan from T.~Gentile of NIST. The single 1\,torr, 4.7\,T result from Nikiel~\cite{nikiel2} is shown for reference.}
	\label{fig:high}
\end{figure}

The top of Figure~\ref{fig:high} shows our achieved, steady-state nuclear polarizations at various $T_D$ relaxation times. Here the relaxation time is largely a function of discharge intensity, with the dimmest discharges resulting in the longest relaxation times. All these measurements were performed on two 1\,torr sealed cells, pumping the $f_4^+$ line with 1.5\,W of laser power for the 1\,T measurements and 3\,W for the others (with the exception of the 2\,T, 449\,s relaxation time point, which was taken at 2\,W). As has been noted by others~\cite{abboud,nikiel}, we do not see a strong steady-state polarization dependence on laser power in the 1 to 4\,W range, and more power than this tends to be counter-productive. 

With the exception of the 1\,T setting, we do not see a strong dependence on the magnetic field. At 2, 3 and 4\,T, polarizations exceeding 80\% were seen, even with relatively short relaxation times below around 300\,s. The comparatively poor performance at 1\,T is likely due to our inability to resolve the Doppler broadened $f_4^+$ and $f_2^+$ lines with the relatively wide frequency band of the pumping laser. With increasing field, the greater separation of these lines made it possible to pump $f_4^+$ without also driving the depolarizing $f_2^+$ transitions. Tuning the pumping laser just to the side of the $f_4^+$ peak, away from $f_2^+$, resulted in higher polarizations, an effect noted in reference~\cite{abboud-thesis}. 

The ability to control the intensity and spatial extent of the discharge plasma proved to be the strongest factor toward attaining the highest polarization. As the magnetic field increases, maintaining and even lighting the discharge plasma becomes more difficult, and the plasma tends to remain closer to the walls of the cell, near the electrodes.  We found that careful adjustment of the frequency and tune of the RF could change the spatial distribution of the plasma inside the cell. The most success was seen when the discharge was as dim as possible while still covering the region where the pumping laser is incident. Our longest relaxation time of 664\,s required a very dim discharge and resulted in a steady-state polarization of 89\%. Our newly-produced sealed cell was able to sustain less intense discharges, providing longer relaxation times, than the older cell on loan from NIST, but aside from the age of the cell, it is not clear why this was the case.

The comparison between these results and the single 32\% measurement at 1\,torr and 4.7\,T from Nikiel \textit{et al}~\cite{nikiel2} appears stark at first glance. However, the noted poor plasma conditions, and the resulting fast relaxation rate, put that result close to our results with similar relaxation times. The Nikiel results cover cell pressures from 1 up to 200 torr, with the achievement of high pumping rates at high pressures being the primary focus.

The bottom of Figure~\ref{fig:high} shows observed pumping rates at each relaxation time where polarization build-up data was taken. Here we follow the convention of Gentile and McKeown~\cite{gentilemck}, where the pumping rate $R_P$ is related to the number of atoms in the cell $N$, and the build-up time and final steady-state polarization extracted from an exponential fit to the polarization build-up $M(t)$: 
\begin{linenomath}\[M(t) = M_{ss}(1-e^{-t/T_P}) \quad \textrm{and} \quad R_P = N M_{ss} / T_P.\] \end{linenomath}
Some build-up sequences were not measured over the full time range prescribed in reference \cite{gentilemck}, so these pumping rate data are provided as rough  guide. Despite this, a trend of decreasing pumping rate with increasing magnetic field is apparent. We have again included the 1\,torr point from Nikiel~\cite{nikiel2} assuming a build-up rate of $0.02/s$ from their given range of rates. Our observed rates are as much as two orders of magnitude lower than typical results at low magnetic field~\cite{gentilemck}, however, even the lowest pumping rate seen ($1.62\times10^{16}$ atoms/s) is more than sufficient to provide the roughly $1.5\times10^{13}$ polarized $^3$He per second to EBIS needed for our source.

\subsection{Discussion of Uncertainty}
Systematic uncertainty in the polarization measurement is limited by using phase-sensitive detection of the plasma absorption and by leveraging the ratios of peak heights. A probe-peak signal is the 1\,kHz component of the voltage measured in the photodiode, which isolates the portion of the laser intensity that is removed by the discharge plasma. Any systematic intensity change across the frequency sweep would occur in both the zero and enhanced polarization signals, canceling its effect on the measurement. The peaks we observe are clean enough to fit well with two Gaussian functions, with typical fitting error in the amplitudes of less than 1\%. These peaks are for the most part free from the background seen from the collisional broadening at higher pressure~\cite{collier}. Because we take roughly 6 seconds to sweep over both peaks, rapid changes in polarization, such as at the beginning of pumping seen in Figure~\ref{fig:relax}, will not be represented completely accurately as one peak is sampled while the other continues to change. This effect can be  corrected via time-interpolation~\cite{abboud-thesis}, however, we are primarily interested in steady-state values.

Optical pumping creates an over-polarization in the $2^3S$ states with respect to spin-temperature equilibrium; this is what drives the build-up of ground-state nuclear polarization. In populations where the pumping rate exceeds the rate at which the ground state is coupled via metastability exchange collisions, the spin-temperature condition does not apply. With increasing magnetic field, the metastability exchanges transfer angular momentum to the ground state less efficiently~\cite{courtade}, as is true for decreasing pressure. Essentially, the high field decoupling of the metastable and ground states means that the probe would measure the pumped, imbalanced populations of the individual excited states \textit{in isolation} from the actual nuclear polarization. The populations of the probed sublevels are affected by this imbalance even though they are not directly addressed by the pump laser, however the \textit{ratio} of these populations are weakly affected by optical pumping~\cite{nikiel2}. This means that the ratio remains nearly equal its spin-temperature value and may still be used for polarization measurements. Were the probe sensitive to the large population changes in the $2^3S$ states during optical pumping, the removal of the pumping laser would show an instantaneous shift in measured polarization when the laser is blocked.  Figure~\ref{fig:relax} shows no discontinuity in the polarization as the pump laser is turned off. As noted in Nikiel \textit{et al.}, even though sharp jumps occur in the individual absorption peaks, the jump is canceled out in their ratio. 

The probe laser itself also acts to drive atoms out of the targeted sublevels, affecting the actual polarization as well as the polarization as measured~\cite{talbot}. The probe laser's total power tends to lower the polarization by emptying the more occupied sublevels. By taking care to reduce the total probe laser power to below 1\,mW, this effect is negligible compared to the effect of the pumping laser. While the total probe power will affect both polarization and build-up with and without the pump light, the per-unit-area intensity can affect the accuracy of the measurement. 
In these measurements, the probe laser intensity was roughly 10\,mW/cm$^2$, significantly higher than the $\sim$0.1\,mW/cm$^2$ of reference~\cite{nikiel2}. To estimate the error expected from such probe intensity, we relied on a simple model of optical pumping and metastability exchange collisions from P.J. Nacher \cite{nacheremail}. This model evaluates pumping rates from given laser intensities assuming a transition linewidth from the radiative lifetime, so while it does not take into account the reduction of the pumping rate from collisional broadening, we can estimate realistic rates by inputting lower intensities. From this model at 4\,T and 1\,torr, a conservative estimate of the error  due to probe intensity in the measured peak ratio at $M=0$ is less than 4\%. A more general model of optical pumping at low pressure is required to more accurately determine the systematic error in these results; future measurements will use lower probe intensity to avoid this complication.

Any inaccuracy in the ``zero polarization'' peak ratio used to calibrate the measurements will introduce error. Residual polarization that has not yet relaxed, or even polarization induced in the plasma by the high field (via polarization of atoms in a magnetized plasma, PAMP~\cite{maul}), could mean that the calibration point is taken with a small positive polarization, rather than at zero.
Error in the peak-ratio measurement of Equation~\ref{eq:pol} propagates as $\sigma_M(r/r_0)=2\sigma_{r/r_0}(r/r_0+1)^{-2}$, so as the ratio of the peaks increases, the effect of the error decreases. This would indicate that even  large errors in the peak ratio at zero polarization have a relatively small effect at high polarization. For example, should the polarization be as high as 5\% when the ``zero'' calibration signal is taken---unlikely even under PAMP---the measured polarization would still be within roughly 1\% of actual when the gas is polarized to near 90\%.

While further study will better quantify the uncertainty in this method, we consider a conservative approximation of the systematic error in the measurements in Figure~\ref{fig:high} to be roughly 4\% absolute.

\section{Conclusion}
Our results represent the highest steady-state polarizations achieved at 1\,torr and above 2\,T to our knowledge. By extending the success of high-field MEOP achieved by the LKB group to low pressure cells, $^3$He polarization exceeding 80\% is attained with relative ease. As we noted with our first observations of this effect~\cite{maxwell2}, sealed cells which we struggled to polarize above 60\% with traditional low field methods were able to reach above 80\% at high field.

Zeeman splitting with increasing magnetic field does act to reduce coupling between electron and nuclear spins to slow the transfer of polarization to the nucleus, however this decoupling also likely inhibits polarization relaxation channels. At high field, the separation of the hyperfine states allow their clear discrimination using a 2\,GHz bandwidth pumping laser, allowing us to cleanly address polarizing transitions and completely saturate those states. The collision broadening of the absorption peaks at higher pressures makes such resolution more difficult.

We expect that the refinement of these techniques could deliver even higher polarization. The use of narrower, longer cells, and more sophisticated electrode schemes could allow a more uniform and easier to maintain plasma discharge. While we used the $f_4^+$ line for optical pumping, the $f_2^-$ line could provide cleaner pumping, particularly at lower field when the  $f_4^+$ and $f_2^+$ lines are difficult to resolve.

With these results, our efforts toward creating a polarized $^3$He ion source for RHIC and a future EIC have been refocused by adapting our scheme to take advantage of high field MEOP. The ability to produce highly polarized, pure $^3$He gas within the EBIS is an unexpected boon, and highlights the value of MEOP techniques even at high magnetic fields.

\subsection*{Acknowledgements}
We gratefully acknowledge Thomas Gentile and P.J. Nacher for their guidance in MEOP techniques. This work was supported by the DOE Office of Nuclear Physics, R\&D for Next Generation Nuclear Physics 
Accelerator Facilities, under grant DE-SC0008740.

\bibliography{high_field_nim}
\bibliographystyle{elsarticle-num}

\end{document}